\input harvmac

\Title{\vbox{\rightline{EFI-97-3}\rightline{hep-th/9703211}}}
{\vbox{\centerline{Matrix Black Holes}}}
\vskip20pt
\centerline{Miao Li~~ {\sl and}~~ Emil Martinec} 
\bigskip
\centerline{\it Enrico Fermi Inst. and Dept. of Physics}
\centerline{\it University of Chicago}
\centerline{\it 5640 S. Ellis Ave., Chicago, IL 60637, USA}

\vskip 1cm
\noindent

Four and five dimensional extremal black holes
with nonzero entropy have simple presentations in M-theory
as gravitational waves bound to configurations of intersecting 
M-branes.  We discuss realizations of these objects
in matrix models of M-theory, investigate the properties
of zero-brane probes, and propose a measure of their
internal density.
A scenario for black hole dynamics is presented.

\Date{3/97}

%
%
\def\p{\partial}
\def\ap{\alpha'}
\def\half{{1\over 2}}

\def\tS{\tilde{S}}

\def\b0{\bar{0}}
\def\b4{\bar{4}}
%
%
%
\def\journal#1&#2(#3){\unskip, \sl #1\ \bf #2 \rm(19#3) }
\def\andjournal#1&#2(#3){\sl #1~\bf #2 \rm (19#3) }

\def\ie{{\it i.e.}}
\def\eg{{\it e.g.}}
\def\cf{{\it c.f.}}

\def\frac#1#2{{#1\over#2}}

\def\half{\frac12}
\def\hf{{\textstyle\half}}

\def\d{\partial}

\def\inbar{\,\vrule height1.5ex width.4pt depth0pt}
\def\IC{\relax\hbox{$\inbar\kern-.3em{\rm C}$}}
\def\IR{\relax{\rm I\kern-.18em R}}
\def\IP{\relax{\rm I\kern-.18em P}}

%
%

\def\npb#1#2#3{Nucl. Phys. {\bf B#1} (#2) #3}

\def\plb#1#2#3{Phys. Lett. {\bf #1B} (#2) #3}
\def\prl#1#2#3{Phys. Rev. Lett. {\bf #1} (#2) #3}

\def\prd#1#2#3{Phys. Rev. {\bf D#1} (#2) #3}

\def\cmp#1#2#3{Comm. Math. Phys. {\bf #1} (#2) #3}
\def\cqg#1#2#3{Class. Quant. Grav. {\bf #1} (#2) #3}

\catcode`\@=11
\def\slash#1{\mathord{\mathpalette\c@ncel{#1}}}
\overfullrule=0pt

\def\VV{{\cal V}}

\def\underrel#1\over#2{\mathrel{\mathop{\kern\z@#1}\limits_{#2}}}

\catcode`\@=12


%

\def \cosh{{\rm cosh}}

\def\exp{{\rm exp}}

\def\bmatrix#1{\left[\matrix{#1}\right]}

\def\fdot{{\dot \varphi}}
\def\xdot{{\dot x}}
\def\rhor{r_{hor}}
%
\nref\bfss{T. Banks, W. Fischler, S.H. Shenker and L. Susskind,
hep-th/9610043.}
\nref\sv{A. Strominger and C. Vafa, hep-th/9601029, 
\plb{379}{1996}{99}.}
\nref\cm{C. Callan and J. Maldacena, hep-th/9602043, \npb{472}{1996}{591}.}
\nref\hs{G. Horowitz and A. Strominger, hep-th/9602051; 
\prl{77}{1996}{2368}.}
\nref\hms{G. Horowitz, J. Maldacena and A. Strominger, hep-th/9603109;
\plb{383}{1996}{151}.}
\nref\cvetic{M. Cvetic and  D. Youm, hep-th/9507090, \prd{53}{1996}{584}.}
\nref\ms{J. Maldacena and A. Strominger, hep-th/9603060; 
\prl{77}{1996}{428}.}
\nref\jkm{C. Johnson, R. Khuri and R. Myers, hep-th/9603061,
\plb{378}{1996}{78}.}
\nref\hlm{G. Horowitz, D. Lowe and J. Maldacena, hep-th/9603195,
\prl{77}{1996}{430}.}
\nref\stretchor{L. Susskind, L. Thorlacius, and J. Uglum,
hep-th/9306069, \prd{48}{1993}{3743}.}
\nref\thooft{G. 't Hooft, unpublished, as cited in \stretchor.}
\nref\dvvcomp{K. Schoutens, E. Verlinde, and H. Verlinde, hep-th/9401081;
Y.Kim, E. Verlinde, and H. Verlinde, hep-th/9502074, \prd{52}{1995}{7053}.}
\nref\rt{J.G. Russo and A.A. Tseytlin, hep-th/9611047.}
\nref\polhem{G. Polhemus, hep-th/9612130.}
\nref\tseytlin{A.A. Tseytlin, hep-th/9604035, \npb{475}{1996}{149}.}
\nref\hormar{G. Horowitz and D. Marolf, hep-th/9605224,
\prd{55}{1997}{835}; hep-th/9610171, \prd{55}{1997}{3654}.}
\nref\callan{C. Callan, J. Maldacena, and A. Peet, hep-th/9510134,
\npb{475}{1996}{645}.}
\nref\dghw{A. Dabholkar, J. Gauntlett, J. Harvey, and D. Waldram,
hep-th/9511053, \npb{474}{1996}{85}.}
\nref\kt{I.R. Klebanov and A.A. Tseytlin, hep-th/9604166,
\npb{475}{1996}{179}.}
\nref\dps{M.R. Douglas, J. Polchinski and A. Strominger, hep-th/9703031.}
\nref\tseytrev{A.A. Tseytlin, hep-th/9702163.}
\nref\dm{S. Das and S. Mathur, hep-th/9606185, \npb{478}{1996}{561}.}
\nref\msi{J. Maldacena and A. Strominger, hep-th/9609026,
\prd{55}{1997}{861}.}
\nref\cgkt{C.G. Callan Jr., S.S. Gubser, I.R. Klebanov and
A.A. Tseytlin, hep-th/9610172.}
\nref\gk{S.S. Gubser and I.R. Klebanov, hep-th/9609076,
\prl{77}{1996}{4491}.}
\nref\dkps{M. Douglas, D. Kabat, P. Pouliot, and S.H. Shenker,
hep-th/9608024, \npb{485}{1997}{85}.}
\nref\jackson{J.D. Jackson, {\it Classical Electrodynamics},
2nd Ed., John Wiley and Sons (1975).}
\nref\jm{J. Maldacena, hep-th/9607235.}
\nref\grt{O. Ganor, S. Rangoolam, and W. Taylor, hep-th/9611202.}
\nref\causality{E. Martinec, hep-th/9304037, \cqg{10}{1993}{L187};
hep-th/9311129.}
\nref\suss{W. Fischler, E. Halyo, A. Rajaraman, and L. Susskind,
hep-th/9703102.}
\nref\taylor{W. Taylor, hep-th/9611042.}
\nref\dvv{R. Dijkgraaf, E. Verlinde, and H. Verlinde, hep-th/9703030.}
\nref\motl{L. Motl, hep-th/9701025.}
\nref\bankseib{T. Banks and N. Seiberg, hep-th/9702187.}
\nref\howu{P.-M. Ho and Y.-S. Wu, hep-th/9703016.}
\nref\imam{Y. Imamura, hep-th/9703077.}
\nref\susslorentz{L. Susskind, hep-th/9308139, \prd{49}{1994}{6606}.}
\nref\dewit{B. de Wit, U. Marquard, and H. Nicolai, \cmp{128}{1990}{39}.}
%

\newsec{Introduction}

D-brane physics, and particular its embodiment in the matrix
model of M-theory \bfss, leads to a radical change in our picture of
gravitational physics.  The static gravitational field is replaced
by a gas of virtual open strings\foot{Or more precisely, their
residual effects at strong coupling, which are the off-diagonal
elements of the matrix.} in a globally flat
background Minkowski spacetime.  Gravitational interactions
turn off below a Planckian distance scale and are replaced by gauge
field dynamics.  In the light of this new perspective,
it is important to revisit the major issues in gravitational
physics.  

In this article, we undertake a preliminary investigation
of black hole dynamics in the context of matrix theory.  
We shall examine two well-known classes of
black holes, the 5D black holes discussed in \sv-\hms, and the
4D black holes discussed in \cvetic-\hlm; section 2 contains
a summary of the pertinent details of their geometry. 
U-duality transformations enable a uniform presentation of these
objects as collections of membranes and fivebranes of M-theory
intersecting along a common string, with the intersection string
carrying a gravitational wave profile.  These configurations
have natural matrix model realizations, which we discuss in
section 5.  The natural probes of the geometry are the
D0-brane/supergraviton `partons' of the matrix formulation.
The trajectories of these probes are by definition the 
light cones of the geometry as seen by low-energy observers
(modulo the effects of the spin connection).
We examine some of the properties of these probes in the
black hole background in section 3.  
In many respects, the background geometry can be thought of
as a kind of optical medium with spatially dependent refractive
index, which is generated by integrating out heavy degrees 
of freedom.  This optical analogy is a central theme of our work.

Section 4 introduces
an intriguing quantity that may be a measure of the density
of matter making up the black hole.  It is the `volume' of
the black hole (in a nonrigorous sense to be explained below)
in the directions transverse to the intersection strings.
This quantity miraculously turns out to be independent
of the moduli of the toroidal compactification, and is just
the number of intersection strings times a factor of order one
in Planck units.  Thus it may be nearly as universal as the
entropy (although it is not U-duality invariant).

In section 5, we present a scenario for black hole dynamics in
the matrix formulation of M-theory.  Because gravitational
dynamics is embedded in a richer structure of noncommutative
variables, there is a natural means to resolve the black hole
evaporation problem by a version of `black hole complementarity'
\stretchor-\dvvcomp.  Moreover, the global
coordinates provided by the infinite momentum frame (IMF)
of the matrix model should enable one to track the zero
brane probes as they become stuck on a `stretched horizon'
\stretchor\ and reradiate as Hawking particles.  The fact
that gravity turns off at short distances, becoming noncommutative
Yang-Mills dynamics, also provides an elegant means of sustaining
the matter making up the black hole against collapse to infinite 
density.

\newsec{Review of 5d and 4d black holes}

The popular version of five dimensional black holes 
discussed in \sv-\hms\ is framed in the IIB theory. 
Compactify the IIB theory on $T^4\times S^1$,
with the $S^1$ having coordinate $x_{5}$ 
and the $T^4$ having coordinates $x_6,...,x_9$.
There are three quantum numbers, $N_1$, $N_5$ and $n_R$, 
corresponding to the number of D1-branes wrapped around $S^1$,
the number of D5-branes wrapped around $T^4\times S^1$ and
the right-moving  momentum along $S^1$, respectively. 
We will often employ a notation 
$$\bmatrix{5&6&7&8&9\cr5&.&.&.&.\cr p_5&.&.&.&.}$$
to denote a set of brane orientations ($p$ denotes momentum along the 
corresponding direction).  
This is an extremal black hole with nonvanishing horizon area. The general
nonextremal black hole is obtained by adding $N_{\bar{1}}$ anti-D1-branes,
$N_{\bar{5}}$ anti-D5-branes and $n_L$ left-moving momentum.
We refer to \hms\ for more details
(see also \rt); in particular, the string metric is
\eqn\fiveIIB{
  ds_{10B}^2=H_5^{1/2}H_1^{1/2}[H_1^{-1}H_5^{-1}(dudv+V du^2)
	+H_5^{-1}(dx_6^2+...+dx_9^2)+ (dx_1^2+...+dx_4^2)]\ .
}

T-dualizing along the 5 direction, one passes to the IIA theory.
All D1-branes are mapped to D0-branes, D5-branes
are mapped to D-4 branes, and the right-moving momentum modes
become right-winding modes along $\tS^1$ and the left-moving momentum
modes become left-winding modes. 
The configuration is now 
$$\bmatrix{.&6&7&8&9\cr 
	   .&.&.&.&.\cr 
	 w_5&.&.&.&.}\,$$ 
with the D0-branes strung along
the winding strings \refs{\hms,\polhem}.
We shall use $N_0 (N_{\bar{0}})$, $N_4 (N_{\bar{4}})$, $w_R (w_L)$
to denote numbers of corresponding branes.   
The strong coupling limit yields M-theory on a circle
of radius $R$ (coordinate $x_{11}$), with configuration
$$\bmatrix{.&6&7&8&9&11    \cr 
	   .&.&.&.&.&p_{11}\cr
	   5&.&.&.&.&11    \cr }\ .$$
An M-theory metric for this situation is known \tseytlin:
\eqn\fiveM{
  ds_{11}^2=F^{2/3}T^{1/3}[F^{-1}T^{-1}(dudv + K du^2)
	+T^{-1} dx_5^2 
	+F^{-1}(dx_6^2+...+dx_9^2)  
	+ (dx_1^2+...+dx_4^2)]\ .
}
Here $F=1+\frac{Q_5}{r^2}$, $T=1+\frac{Q_2}{r^2}$,
$K=\frac{P}{r^2}$ are harmonic functions, with $Q_5$, $Q_2$, and
$P$ the fivebrane, membrane, and `longitudinal wave' momentum.

Horowitz and Marolf \hormar\ have considered the geometry of the
six-dimensional black string that results from decompactifaction
(or large radius)
of the longitudinal direction $x_{11}=u+v$, in the particular
case $F=T$.  They note a generalization \refs{\callan,\dghw}
of \fiveM\ to include travelling waves with 
both `longitudinal' and transverse polarizations:
\eqn\transverse{\eqalign{
  ds^2=\left(1+\frac{r_*^2}{r^2}\right)^{-1}& \left[dudv+
	\frac{p(u)+r_*^2\fdot^2(u)}{r^2} du^2\right]
	+\left(1+\frac{r_*^2}{r^2}\right)(dr^2+r^2d\Omega_3^2)\cr
		&-\frac{2r_*^2}{r_*^2+r^2}\sum_{i=6}^9\fdot_i(u) dx^i du 
		+(dx_5^2+...+dx_9^2)\ .
}}
It is consistent to interpret the `longitudinal waves' 
with a coarse-graining of the transverse waves (which are all one sees in
D-brane physics) below some wavelength cutoff.
The general features of the geometry are captured by keeping
only the `longitudinal' waves, and for the remainder of
this section we shall do so, setting $\fdot_i(u)=0$; furthermore
let $p(u)=const.=r_*^4\sigma^2$.

The horizon at $r=0$ is a nonsingular null surface, as may be seen
by passing to coordinates 
\eqn\newcoords{\eqalign{
  U=~&\frac{1}{2\sigma} e^{2\sigma u}\cr
  V=~&v-{\hat R}^2\sigma\cr
  W=~&e^{-\sigma u}{\hat R}^{-1}\ .
}}
where ${\hat R}=r_*[\frac{(r_*^2+r^2)}{r^2}]^{1/2}$.
In these coordinates, the horizon at $U=0$ looks like 
$adS_3\times S^3\times\IR^5$:
\eqn\horgeom{
  ds^2\sim r_*^2\bigl(-W^2dUdV+(d\log W)^2+d\Omega_3^2\bigr)
	+(dx_5^2+...+dx_9^2)\ .
}
The radius of curvature at the horizon, $r_*$, 
is quite large for large black holes, and the horizon geometry is
smooth \hormar.  The interior geometry
is of the same form as \fiveM, with the replacements
$1+\frac{Q_i}{r^2}\rightarrow -1+\frac{Q_i}{r^2}$, etc.
Timelike geodesics reach the horizon in finite proper time, and
null geodesics reach the horizon at finite values of their
affine parameter.  The continuation of probe motion beyond the
horizon (ignoring back-reaction, etc.) sees the probe reach
a minimum radius, bounce, and hit a Cauchy horizon \hormar.

The four dimensional extremal black hole has a similar M-theory
interpretation as a configuration
$$\bmatrix{4&5&6&7&.&.&11\cr 
	   .&.&6&7&8&9&11\cr 
	   4&5&.&.&8&9&11\cr
	   .&.&.&.&.&.&p_{11}}$$
of intersecting fivebranes.  A metric involving longitudinal waves
is \refs{\tseytlin,\kt}
\eqn\fourM{\eqalign{
  ds_{11}^2=&(F_1F_2F_3)^{2/3}[(F_1F_2F_3)^{-1}(dudv+K du^2)
	+(dx_1^2+...+dx_3^2)\cr
	&+(F_3F_1)^{-1}(dx_4^2+dx_5^2)
        +(F_1F_2)^{-1}(dx_6^2+dx_7^2)
        +(F_2F_3)^{-1}(dx_8^2+dx_9^2)]\ .
}}


\newsec{Geometry and probes}

The geometry of the black holes \fiveM, \fourM\ 
is measured in different ways by various probes.
We will mostly be interested in 0-brane probes,
which can be statically supported by a BPS cancellation
of gravitational and gauge forces.
The cancellation is spoiled when the black hole and probe
have relative velocity, causing the probe to be attracted to
the hole.
The Born-Infeld action for D0-branes is a simple consequence
of massless particle dynamics in eleven dimensions.
Begin with the massless particle action in 11d:
\eqn\massless{
  S=\int p_M\xdot^M-\hf eG^{MN}p_Mp_N\ ,
}
where $M=0,1,...,9,11$ are 11d coordinates;
10d labels will be $\mu=0,1,...,9$.  Take the metric to
have the Kaluza-Klein form
\eqn\Mthmetric{
  ds^2=e^{-2\phi/3}g_{\mu\nu}dx^\mu dx^\nu
	+e^{4\phi/3}(dx_{11}-A_\mu dx^\mu)^2\ ;
}
eliminating $p_\mu$ (note that it is the inverse metric which appears
in \massless), one finds
\eqn\IIAmassless{
  S=\int \frac1{2e}e^{-2\phi/3}g_{\mu\nu}\xdot^\mu\xdot^\nu
	-p_{11}A_\mu\xdot^\mu-\frac e2 e^{-4\phi/3}p_{11}^2
		+p_{11}\xdot_{11}\ .
}
Finally, solving for the einbein $e$ yields the D0-brane action
\eqn\dzerobrane{
  S=\int p_{11}[e^{-\phi}\sqrt{g_{\mu\nu}\xdot^\mu\xdot^\nu}
	-A_\mu\xdot^\mu+\xdot_{11}]\ .
}
The last term is a total derivative when $p_{11}$ is constant,
but contributes to the eikonal phase of the particle.
The inclusion of the fermionic terms in the D0-brane action
simply generates the coupling of the background fields to
the intrinsic spin of the 11d supergraviton multiplet.

\subsec{5D black holes}

The action for a nonrelativistic D1-brane probe derived in 
\refs{\dps,\tseytrev} can be recast as the D0-brane action
\eqn\nonr{S=-{1\over R}\int d\tau +{1\over 2R}\int d\tau\left(
FTv^2 +T w^2\right)\ ,}
where $R$ is the radius of the eleventh dimension and 
$$\eqalign{
T &= 1+{r_w^2\over r^2}\ , 
	\quad r_w^2= {l_p^6w_R\over (2\pi)^4V}\ , \cr
F &=1+{r_4^2\over r^2}\ , \quad 
	r_4^2={l_p^3N_4\over (2\pi)^2 R_5}\ .}$$
Here $v$ is the velocity in the macroscopic dimensions, while
$w$ is that in the internal torus; we shall set $w=0$. 
It was shown by \dps\ that a nonrelativistic
zero-brane probe is captured if
its impact parameter is less than $r_w+r_4$.  Note that this
is independent of the zero-brane charge $Q$ carried by the black hole.

The properties of this probe can be recast in eleven-dimensional
form as follows:
Let us assume that the action \nonr\ results from
an eikonal equation for the 11d supergraviton of the form
\eqn\eik{n^2(E^2-p_{11}^2)=p_i^2\ ,}
where $p_i=\p_i \psi$, $p_{11}=\p_{11}\psi$ and $E=\p_t\psi$;
$\psi$ is the eikonal of the wave function,
$\Phi\sim e^{i\psi}$, and $n(r)$ 
is a spatially dependent `refractive index'. 
Let $p_+=E+p_{11}$ and $p_-=E-p_{11}$;
both of these quantities are conserved in 
the black hole background \fiveM.\foot{Although the generic metric
for a wave profile \transverse\ only preserves the null
Killing vector $p_{-}=\d/\d v$ corresponding to light-cone time.}
The light-cone energy is
$$p_+={p_i^2\over n^2p_-}={R p_i^2\over 2n^2}\ ,$$
where we used the approximation $p_+=2/R$. This implies the action
\eqn\eac{S={1\over 2R}\int d\tau n^2v^2\ ,}
compared with \nonr\ we find $n^2=FT$.  
Thus, for the super-graviton 
propagating in the background of the 11D black hole, we have the 
eikonal equation
\eqn\eikn{n^2\left((\p_t\psi)^2-(\p_{11}\psi)^2\right)
=(\p_i\psi)^2\ ,}
with $n^2=FT$. 

Indeed, the massless particle action \massless\ in the background
\fiveM\ yields the scalar Laplacian
\eqn\lapfive{
  FT(\d_u\d_v- K\d_v^2)+(\d_1^2+...+\d_4^2)
	+T\d_5^2+F(\d_6^2+...+\d_9^2)
}
Inclusion of fermionic terms in \massless\ will generate
the spin connection terms of the higher spin wave equations obeyed
by supergravitons.  
It is possible to choose polarizations such that these terms
vanish (for example the $A_{567}$ component of the antisymmetric
tensor field).
This Laplacian reduces to \eikn\
for waves with $\d_u\psi\gg\d_v\psi$,
and no dependence on internal coordinates,
unless one is close to the horizon\foot{One can arrange a hierarchy
of scales so that the region where the $\d_v^2$ term
becomes important is much inside $r_w$, $r_4$.}.
Thus the corresponding wave equation for a D0-brane at low energies is
\eqn\lap{n^2(\p_t^2-\p_{11}^2)\Phi=\p_i^2\Phi\ .}
Consider S-wave scattering of the D0-brane off the black hole.
Since both energy and $p_{11}$ are conserved, we can replace the
l.h.s. of \lap\ by $-n^2(\omega^2-p_{11}^2)\Phi = -n^2\omega'^2\Phi$.
The relevant wave equation is then 
\eqn\laps{r^{-3}\p_r (r^3\p_r\Phi)+\omega'^2n^2\Phi =0.}
The low energy limit requires 
$\omega' r_w, \omega' r_4 \ll 1$. 
As in \dm\ and \msi, we shall solve this equation approximately in
two regions. Region I: $r\gg\omega' r_wr_4$. 
Region II: $[Q(\frac{\omega-p_{11}}{\omega+p_{11}})]^{1/2}\ll r\ll r_w, r_4$.
The low energy limit ensures that there is an 
overlap between the two regions. 
The region inside of region II does not substantially affect
the results, as one may see by comparing our answer below
with a similar calculation of \msi,
which finds the same result for low frequency waves
by a more involved computation using the exact Laplacian \lapfive.

In region I, the wave equation is approximately
\eqn\lapi{
  r^{-3}\p_r (r^3\p_r\Phi)+\omega'^2\Phi=0\ ,
}
the general solution is
$$\Phi_I= \sqrt{\omega' r}\left( \alpha J_1(\omega' r)+\beta N_1(
\omega' r)\right).$$
In region II, the equation reduces to
\eqn\lapii{
  r\p_r (r^3\p_r\Phi)+\omega'^2r_w^2r_4^2\Phi =0\ .
}
Letting $\rho=1/r$ and $\Phi =\rho\Psi$, we have an equation
\eqn\lapiii{
  \p_\rho^2\Psi +{1\over\rho}\p_\rho\Psi+
	(\omega'^2r_w^2r_4^2-{1\over \rho^2})\Psi=0\ ,
}
whose solution is again given by Bessel functions. 
If we demand that there is only incoming wave near the horizon
($\rho=\infty$), we have to choose 
$\Psi =J_1(\omega'r_wr_4\rho) -iN_1(\omega'r_wr_4\rho)$. 
So in region II the approximate solution is
$$\Phi_{II}=A\rho\left(J_1(\omega'r_wr_4\rho)
  -iN_1(\omega'r_wr_4\rho)\right)\ .$$
Matching the two solutions in the overlapping region, one finds
that $|\alpha|\gg|\beta|$ as usual, and
\eqn\match{\alpha ={4iA\over \pi\omega'^{5/2}r_wr_4}\ .}

The incoming flux at $r=\infty$ is given by
$$f_{in}={1\over 2i}(\Phi_{in}^*r^3\p_r\Phi_{in}-c.c.)
  ={\omega'\over 2\pi}|\alpha|^2\ ,$$
and the absorption flux at the horizon is
$$f_{abs}={1\over 2i}(\Phi^*\rho^{-1}\p_\rho \Phi -c.c.)
  ={2\over \pi}|A|^2\ .$$
Thus the absorption ratio is
\eqn\absr{
  \sigma^S_{abs}={f_{abs}\over f_{in}}={1\over 4}
	\pi^2\omega'^4r_w^2r_4^2\ ,
}
and the absorption cross section
\eqn\absc{\sigma_{abs}={4\pi\over \omega'^3}\sigma^S_{abs}
=\pi^3\omega'r_w^2r_4^2\ .}
This is to be contrasted to the results for a minimally coupled 
scalar \msi, and for a fixed scalar \cgkt. For a minimally coupled
scalar, in the low energy limit the absorption cross section is
independent of $\omega$ but proportional to the horizon area. For
a fixed scalar, the absorption cross section goes as $\omega^2$
in the low energy limit.  
We will discuss the interpretation of the
result \absc\ in section 4.

\subsec{4D black holes from intersecting 5-branes}

Consider three sets of 5-branes intersecting along a string
as in \fourM, and take
the string as the longitudinal direction. Let the compact space 
transverse to the string be $T^6$, 
each of whose circles has size $L$. 
The 11D Einstein metric is \fourM\ with
\eqn\determ{\eqalign{
  F_i&=1+{r_i\over r}, 
	\quad r_i={l_p^3N_i\over 2L^2},\cr
	K &=Q/r, \quad Q\sim N_0,
}}
where $N_i$ are numbers of 5-branes, and $N_0$ the number of D0-branes.
It is easy to read off the relevant quantities from \fourM:
\eqn\reda{\eqalign{
  e^{-\phi}& =(F_1F_2F_3)^{1/4}(1+K)^{3/4}, \quad
	G_{00}=((1+K) F_1F_2F_3)^{-1/2},\cr
	G_{ij}& =\delta_{ij}((1+K)F_1F_2F_3)^{1/2}, \quad A_0=(1+K)^{-1}-1,
}}
where $G_{\mu\nu}$ is the string metric.

The action of a probing nonrelativistic D0-brane can be derived from 
the Dirac-Born-Infeld action \dzerobrane.
Again, just like \nonr, there is no static potential
\eqn\annon{S=-{1\over R}\int d\tau +{1\over 2R}\int d\tau F_1F_2F_3v^2.}
Consequently, the wave equation is the same as \eikn\ with
$n^2=F_1F_2F_3$.

The approximate wave equation
\eqn\lapiv{
  n^2(\omega^2-p_{11}^2)\Phi +r^{-2}\p_r(r^2\p_r\Phi)=0
}
for the S-wave is identical to the one treated in \gk\ in the limit
$r_0=0$. 
Again this is an approximation 
(for wavefunctions with $\d_u\Phi\gg\d_v\Phi$) 
to the exact wave equation, which is
\eqn\lapfour{
  F_1F_2F_3(\d_u\d_v- K\d_v^2)+(\d_1^2+...+\d_3^2)
	+F_3F_1(\d_4^2+\d_5^2)
	+F_1F_2(\d_6^2+\d_7^2)
	+F_2F_3(\d_8^2+\d_9^2)\ .
}
The s-wave absorption cross section in the low energy limit 
of the 4d black hole is
\eqn\crosiv{
  \sigma_{abs}=4\pi^2r_1r_2r_3\omega'\ ,
}
where again $\omega'^2=\omega^2-p_{11}^2$.  
The above scalar wave equation as well as \lapfive\
are straightforward consequences of the `harmonic function rule'
\tseytlin; for modes independent of internal coordinates,
they both take the form
  $$\Delta=n^2(\d_u\d_v- K\d_v^2)+\d_i^2\ .$$

Douglas, Polchinski, and Strominger \dps\ have reproduced the
probe action \nonr\ up to terms of order $1/r^2$ in a D-brane 
calculation.  The gravitational field of the black hole is 
generated by the exchange of closed strings between the D-branes;
in the dual open string channel, these are virtual loops
of open strings whose mass is $r$ in string units.  
Thus in a very direct sense the background geometry {\it is}
a spatially dependent permeability contributed by 
the vacuum polarization.  A profound
feature of this picture is that a curved space geometry is
generated by virtual effects of objects in flat space.
Note also
that the mass of the strings being integrated out is determined
by their length in flat space, not the spacelike distance to
the horizon in the black hole metric (which is infinite, of course).
There appear to be some subtleties \dps\ in reproducing the $1/r^4$
terms in \nonr.

\newsec{The `transverse volume' of a black hole}

In this section we will view the above black holes as black strings
wrapped around $x^{11}$ in M-theory. The area
of the horizon is then nine-dimensional 
including the longitudinal direction.
Let $A^E_9$ denote the area of the 9D horizon, measured against
the 11D Einstein metric; it is not difficult to show that
the proper definition of the entropy is
\eqn\ent{S={A^E_9\over \pi l_p^9}\ .}
Here $l_p$ is the 11D Planck length, in terms of which the membrane tension
is given by $T_2=l_p^{-3}$. 
If more dimensions are compactified,
the formula \ent\ can be re-written as $A_{D-2}/(4G_D)$, where
$A_{D-2}$ is the horizon area of the D dimensional black hole, and
$G_D$ is the D dimensional Newton constant.

While the entropy is a pure number independent of both $R$, the radius
of the longitudinal dimension, and $l_p$, one does not expect
it be invariant when the whole system is boosted along the 
longitudinal direction.  Indeed, the longitudinal momentum
is proportional to $N_0-N_{\bar{0}}$, the difference of the
number of 0-branes and the number of anti-0-branes, and $S$
in all cases is proportional to $\sqrt{N_0}+\sqrt{N_{\bar{0}}}$.
On the other hand, there is an intriguing 
`geometric' quantity that {\it is} invariant
under longitudinal boosts:
\eqn\cross{\Sigma ={V^E_9\over \pi l_p^9}\ ,}
where $V^E_9$ is the transverse `volume' enclosed by the horizon,
in some sense measured against the 11D Einstein metric. 
We define it as the $r\rightarrow r_0$ limit ($r_0$ is the
horizon radius)
of the volume of a {\it Euclidean} (D-2)-sphere
of radius $r$, times the part of the eleven-dimensional
volume element $\sqrt{G_\perp}$ transverse to the
eleventh dimension.
Roughly speaking, $V^E_9$ is proportional to 
the optical transverse cross section (see below).
According to the
standard infinite momentum frame physics, this quantity is
boost invariant, and therefore will have a simple description
in the matrix model. 
We shall see that, rather surprisingly, the
ratio \cross, not only is independent of $N_0$ and $N_{\bar{0}}$
but also independent of $R$ and $l_p$ just like the entropy. 
Moreover, it has a simple dependence on numbers of other types of branes.
In the extremal limit, it is linear in the number of branes of
any type other than D0-branes. 
The definition of $\Sigma$ in
\cross\ is such that in cases examined here it is always a rational
number for extremal black holes.
Given these properties, clearly it is an important quantity
to study in addition to the entropy.
On the downside, $V^E_9$ is not U-duality invariant, selecting
as it does 0-branes for special treatment.

The relation between the 11D Einstein metric and 
the IIA string metric is 
  $$G_{\mu\nu}^E=e^{-2\phi/3}G^s_{\mu\nu}\ ,\qquad
	\mu=0,1,...,9\ .$$
Our convention for the dilaton is such that its asymptotic
value is always zero, so the effective string coupling constant
is $g=g_se^\phi$, $g_s$ is the asymptotic value of the string coupling
constant. The transverse volume viewed in terms of the IIA theory
is just the spatial volume enclosed by the horizon. Thus
$$V^E_9=e^{-3\phi}V^s_9,$$
where $V_9^s$ is the volume measured in the string metric.

We assume the
geometry of the horizon is always the tensor product of a $D-2$ dimensional 
sphere and some compactified space of dimension $10-D$, so the
black hole is really a hole in $D$ dimensional spacetime. 
We ignore the curved space geometry
and take as a measure of the volume enclosed by the horizon
that of the standard Euclidean ball enclosed by the $D-2$ sphere
times the volume of the compact space. 
This is not completely unreasonable, 
given the way that D-branes (and matrix theory)
reproduce curved space geometry from Euclidean matrix dynamics
(see \refs{\dkps,\dps} and below).
Let $A^s_8$ be the horizon
area viewed in 10D, then $V^s_9={1\over D-1}A^s_8\rhor$; $\rhor$ is the
radius of the horizon.  We thus have
\eqn\prel{
	\Sigma ={1\over (D-1)\pi l_p^9}e^{-3\phi}4G(r_0)S\;\rhor\ ,
}
where we used the formula $S=A_8^s/(4G(r_0))$, $G(r_0)$ is the 10D Newton
constant at the horizon. Its relation with $\phi$ is
$G(r_0)=8\pi^6(\ap)^4g_s^2e^{2\phi}$. Substituting this relation into
\prel,
\eqn\rel{
  \Sigma={32\pi^6\over (D-1)\pi l^9_p}(\ap)^4g_s^2Se^{-\phi}\rhor\ .
}
This is a rather obscure relation. Now we make a simple
observation about the condition for $\Sigma$ to be a number independent
of $l_p$ and $R$.  As is to be seen, 
for the black holes we are considering, $e^{-\phi}$ at the horizon
is always a number independent of $l_p$ and $R$. If 
the dependence of $\rhor$ on $g_s$ and $\ap$ is always linear in the
combination $g_s\sqrt{\ap}=R$, then combined with $(\ap)^4g_s^2$
this gives rise to a number
$$(\ap)^4g_s^2g_s\sqrt{\ap}=(\ap)^3R^3= {l_p^9\over (2\pi)^6},$$
where we have used the relation $\ap=l^3_p/(4\pi^2R)$. This together
with \rel\ implies that $\Sigma$ is a number independent of $l_p$ and 
$R$. For the time being, we have no general argument 
in terms of the usual string theory for why $\rhor$ is always
proportional to $g_s\sqrt{\ap}$, let alone the fact that $\Sigma$
is always independent of the 0-brane charge. In the following we 
shall examine the 5D black holes and the 4D black holes separately.

\bigskip
\subsec{5D Black Holes}

Consider again an extremal black hole \fiveIIB\ of
the IIB theory on $T^4\times S^1$.
This is an extremal black hole with nonvanishing horizon area. 
The general nonextremal black hole is obtained 
by adding $N_{\bar{1}}$ anti-D1-branes,
$N_{\bar{5}}$ anti-D5-branes and $n_L$ left-moving momentum.
We refer to \hms\ for formulas for the string metric and more details.
For our purposes, we need to know the dilaton, the 
space dependent radius $R_5$ of $S^1$ and the horizon radius $\rhor$:
\eqn\data{\eqalign{
e^{2\phi}& =f_0f_4^{-1}\cr
R_5(r_0)& =f_0^{-1/4}f_4^{-1/4}f_w^{1/2}R_5(\infty)\cr
\rhor& =f_0^{1/4}f_4^{1/4}r_0\ ,
}}
where $f_i=\cosh\alpha_i$. All scales are measured in the string metric.

Let us T-dualize along the $S^1$ to go to the IIA theory.
We still use $R_5(r_0)$ to denote the new radius in the
T-dual theory, and $\phi$ the dilaton in the IIA theory. Using
the standard T-duality transformation we find
\eqn\newf{\eqalign{e^{2\phi}&=f_0^{3/2}f_4^{-1/2}f_w^{-1},\cr
R_5(r_0)&=f_0^{1/4}f_4^{1/4}f_w^{-1/2}R_5(\infty), \cr
\rhor& =f_0^{1/4}f_4^{1/4}r_0
}}

The scale factors
$f_i$ can be read off from relations in \hms
\eqn\frel{\eqalign{f_0^{1/2} &={1\over 2}(N_0N_{\bar{0}})^{-1/4}(\sqrt{N_0}+
\sqrt{N_{\bar{0}}}),\cr
f_4^{1/2} &={1\over 2}(N_4N_{\bar{4}})^{-1/4}(\sqrt{N_4}+
\sqrt{N_{\bar{4}}}),\cr
f_w^{1/2} &={1\over 2} (w_R w_L)^{-1/4}(\sqrt{w_R}+
\sqrt{w_L}).}}
We also need a formula for $r_0$ which is completely determined
in terms of the other parameters:
\eqn\mf{r_0=2g_s\sqrt{\ap}\left({N_4N_{\bar{4}}w_Rw_L\over N_0N_{\bar{0}}}
\right)^{1/4},}
where $g_s$ is the coupling constant for the IIA theory. The above
formula together with the last relation in \data\ is precisely what
we need to get a pure number $\Sigma$. To compute $\Sigma$ using \rel, 
we first compute the combination 
$e^{-\phi}\rhor=f_0^{-1/2}f_4^{1/2}f_w^{1/2}r_0$.
It is quite remarkable that all the factors such as $N_0N_{\bar{0}}$
cancel out, thanks to eq.\mf. Moreover, due to the factor $f_0^{-1/2}$
we obtain a factor $1/(\sqrt{N_0}+\sqrt{N_{\bar{0}}})$ which is to cancel
the factor $(\sqrt{N_0}+\sqrt{N_{\bar{0}}})$ in $S$. 
The final result is
\eqn\fres{\Sigma ={1\over 4}(\sqrt{N_4}+ \sqrt{N_{\bar{4}}})^2
(\sqrt{w_R}+\sqrt{w_L})^2,}
where we used the formula $S=2\pi (\sqrt{N_0}+\sqrt{N_{\bar{0}}})
(\sqrt{N_4}+\sqrt{N_{\bar{4}}})(\sqrt{w_R}+\sqrt{w_L})$. Eq.\fres\ is
a remarkably simple result compared to any geometric datum involved
in the definition of $\Sigma$. As we shall see, there is a similar formula
for 4D black holes.

We have emphasized that the relation $\rhor\sim g_s\sqrt{\ap}=R$ is crucial
for $\Sigma$ to be independent of $l_p$ and $R$. In no way is this relation
a consequence of string theory, since in the IIB picture, the relation
becomes $\rhor\sim \sqrt{g_s\ap}$, where now $g_s$ is the 
IIB string coupling constant.

To show that $\Sigma$ is the only quantity independent of $N_0$, we list
all three scales first in string metric
\eqn\tsca{\eqalign{
  \rhor^s &=f_0^{1/4}f_4^{1/4}r_0=2R\left({f_0f_4N_4N_{\bar{4}}
	  w_Rw_L\over N_0N_{\bar{0}}}\right)^{1/4},\cr
	V^s &={\alpha'}^2f_0f_4^{-1}
	  \left({N_0N_{\bar{0}}\over N_4N_{\bar{4}}}\right)^{1/2}, \cr
	R^s_5 &={\sqrt{\ap}\over g_s}f_0^{1/4}f_4^{1/4}f_w^{-1/2}
	  \left({N_0N_{\bar{0}}\over w_Rw_L}\right)^{1/2},}}
where $(2\pi)^4V^s$ is the volume of $T^4$. We see that all three
scales in string metric depend on $N_0$. In the 11D Einstein metric,
these become
\eqn\tsc{\eqalign{
  \rhor^E &=2Rf_4^{1/3}f_w^{1/6}\left({N_4N_{\bar{4}}
	w_Rw_L\over N_0N_{\bar{0}}}\right)^{1/4}, \cr
	V^E &={\alpha'}^2f_4^{-2/3}f_w^{2/3}
	\left({N_0N_{\bar{0}}\over N_4N_{\bar{4}}}\right)^{1/2}, \cr
	R^E_5 &= {\sqrt{\ap}\over g_s}f_4^{1/3}f_w^{-1/3}
	\left({N_0N_{\bar{0}}\over w_Rw_L}\right)^{1/2}.}}
Now all three scales still depend on $N_0$.  
Interestingly there is no dependence on $f_0$.
These three scales assume a more obscure form
in the 10D Einstein metric.

From \tsca\ and \tsc\ we see that both internal scales expand
with $N_0\rightarrow \infty$, if one holds other numbers fixed. 
The horizon size $\rhor$ contracts in the large $N_0$ limit. 
The extremal limit corresponds to vanishing 
$N_{\bar{0}}$, $N_{\bar{4}}$ and $w_L$. 
In this limit scales in \tsca\ and in \tsc\ become free
parameters, except that their combination $\Sigma$ is fixed as in
\fres. 

In computing $A^E_9$, the quantity associated to entropy \ent, we
trade the radial size $\rhor^E$ in $V^E_9$ with the longitudinal size
$R$. Since $V^E_9$ is independent of $N_0$, one might conclude that
$A^E_9$ is also independent of $N_0$, in contradiction with
the entropy formula. The resolution to this puzzle is obvious. One should 
not use $2\pi R$ in computing $A^E_9$, but the effective longitudinal
size at the horizon: $2\pi R\exp (2\phi/3)=2\pi R f_0^{1/2}f_4^{-1/6}
f_w^{-1/3}$.  This formula shows that in the large $N_0$ limit, the 
longitudinal size grows as $N_0^{1/2}$. In terms of the longitudinal
momentum $p_{11}$, we have $R^E\sim (p_{11})^{1/2}$, an interesting
result to be explained in matrix theory.

The transverse volume defined here has an interesting 
connection to the capture cross-section computed in section 3.
Substituting the formulas for $r_w$ and $r_4$ into
\absc\ and defining a new quantity $(2\pi)^4V2\pi R_5\lambda\sigma_{abs}$,
one finds
\eqn\trans{
  (2\pi)^4V2\pi R_5\lambda\sigma_{abs}={\pi^3}l_p^9w_RN_4\ ,
}
where $\lambda=2\pi/\omega'$.  This is just the transverse volume.
The formula \trans\ also has an optics interpretation.  
In the long wavelength limit, the absorption
cross-section of a dielectric body of volume $\VV$ is
(see for instance \jackson)
\eqn\dielect{
  \sigma_{abs}=\frac{8\pi^2}{\lambda}\cdot\VV\cdot {\rm Im}(4\pi\chi)\ ,
}
where $4\pi\chi$ is the dielectric susceptibility.
Thus $\Sigma$ does indeed play the role of the transverse
volume of the black hole!

Given this interpretation, one expects a similar formula
for the absorption cross section of a nonextremal black hole, in which
instead of the product $w_RN_4$ there is a factor $(\sqrt{w_R}
+\sqrt{w_L})^2(\sqrt{N_4}+\sqrt{N_{\bar{4}}})^2$. Thus the formula
contains no thermal factors as for other scalars (such as
$\exp(\omega/T_H)-1$). This must be the case since temperatures
usually depend on the D0-brane charge carried by the black hole, while
the probing D0-brane decouples from this charge up to order $v^2$.

\bigskip
\subsec{4D black holes}

Compactifying the IIA theory on $T^4\times S^1\times S'^1$,
the 4D black holes considered in \ms\ and \hlm\ in general carry
4 different charges. These are charges associated to D6-branes
wrapped around $T^4\times S^1\times S'^1$, NS5-branes wrapped
around $T^4\times S^1$, D2-branes wrapped around $S^1\times S'^1$,
and string momentum modes flowing around $S^1$. 
The configuration is 
$$\bmatrix{4&5&6&7&8&9\cr 
	   .&{\bf5}&{\bf6}&{\bf7}&{\bf8}&{\bf9}\cr 
	   4&.&.&.&.&9\cr
	   .&.&.&.&.&p_9}\ ;$$
fat characters denote NS fivebranes.
There are 8 independent integers, 4 of them are numbers of branes, 
and 4 are numbers of anti-branes. To put this class of black-holes in
the context of matrix theory, we need to T-dualize along both
$S^1$ and $S'^1$. D6-branes become D-4 branes wrapped around
$T^4$, D2-branes become D0-branes, NS5-branes become the so-called
`non-marginal solitonic branes'
\tseytrev, and finally momentum modes flowing
along $S^1$ become winding modes along $\tS^1$.
The strong coupling limit yields the configuration
$$\bmatrix{.&5&6&7&8&.&11\cr 
	   .&5&6&7&8&9&.\cr 
	   .&.&.&.&.&.&p_{11}\cr
	   .&.&.&.&.&9&11}\ .$$
We can say less about this system in matrix theory
(although \fourM\ provides a similar equivalent system; see below).
The main roadblock is that we do not quite understand 
the transverse 5-brane which
is T-dual of one of the constituents in the black hole. 
However, as
we shall see, there is a formula for $\Sigma$ similar to that
in \fres, thus crying for some simple explanation.

Before the T-duality transformation,
the dilaton and string metric are \jm
\eqn\dm{\eqalign{
  e^{2\phi} &=f_0^{-1/2}f_5^{-1}f_4^{3/2},\cr
  ds_{str}^2 &=f_0^{-1/2}f_4^{-1/2}(-dt^2+dx_9^2+k(dt-dx_9)^2)
	+f_0^{-1/2}f_4^{-1/2}f_5dx_4^2 \cr
	& +f_0^{1/2}f_4^{-1/2}(dx_5^2+\dots +dx_8^2)+f_0^{1/2}f_4^{1/2}f_5
	(dx_1^2+\dots +dx_3^2)\ ,
}}
where the indices of $f_i$ are named after branes in the T-dual theory.
$T^4$ is parametrized by $(x_5, \dots , x_8)$, $S^1$ is parametrized
by $x_9$, $S'^1$ parametrized by $x_4$. Factors $f_i$ are harmonic
functions, and at horizon assume forms similar to those in \frel.
It follows from the metric \dm
\eqn\rad{\eqalign{
  R_4(r_0) &=f_0^{-1/4}f_4^{-1/4}f_5^{1/2}R_4(\infty),\cr
  R_9(r_0) &=f_0^{-1/4}f_4^{-1/4}f_w^{1/2}R_9(\infty).}}
Again we still use $R_4$ and $R_9$ to denote radii of $\tS'^1$ and
$\tS^1$ in the T-dual theory.  These and the new dilaton are
\eqn\nd{\eqalign{
  e^{2\phi} &=f_0^{3/2}f_4^{-1/2}f_w^{-1}, \cr
  R_4(r_0) &=f_0^{1/4}f_4^{1/4}f_5^{-1/2}R_4(\infty),\cr
  R_9(r_0) &=f_0^{1/4}f_4^{1/4}f_w^{-1/2}R_9(\infty).}}
Note that the dilaton has the same form as in the 5D case \newf.
After certain amount of calculation, we find $R_i(\infty)$ and
the horizon size
\eqn\hs{\eqalign{R_4(\infty) &=R\left({N_4N_{\bar{4}}w_Rw_L\over N_0N_{\bar{0}}
N_5N_{\bar{5}}}\right)^{1/4},\cr
R_9(\infty) & ={\ap\over R}\left({ N_0N_{\bar{0}}\over w_Rw_L}
\right)^{1/2}, \cr
\rhor &=2Rf_0^{1/4}f_4^{1/4}f_5^{1/2}\left({N_4N_{\bar{4}}N_5N_{\bar{5}}
w_Rw_L\over  N_0N_{\bar{0}}}\right)^{1/4}.}}
So indeed $\rhor$ scales as $R$ to ensure that $\Sigma$ be independent 
of $R$. We see that three scales $R_4$, $R_9$ and $\rhor$ in the string
metric all depend on $f_0$ therefore on $N_0$. Let $(2\pi)^4V(r_0)$
denote the volume of $T^4$ in string metric, then
\eqn\fs{V(r_0)=f_0f_4^{-1}(\ap)^2\left({N_0N_{\bar{0}}\over 
N_4N_{\bar{4}}}\right)^{1/2},}
also depends on $f_0$. In the 11D Einstein metric, these four scales
are
\eqn\ffs{\eqalign{\rhor^E &=2Rf_4^{1/3}f_5^{1/2}f_w^{1/6}
\left({N_4N_{\bar{4}}N_5N_{\bar{5}}   w_Rw_L\over N_0N_{\bar{0}}
}\right)^{1/4}, \cr
R^E_4 &= f_4^{1/3}f_5^{-1/2}f_w^{1/6}R_4(\infty),\cr
R^E_9 &= f_4^{1/3}f_w^{-1/3}R_9(\infty),\cr
V^E &=f_4^{-2/3}f_w^{2/3}(\ap)^2\left({N_0N_{\bar{0}}\over 
N_4N_{\bar{4}}}\right)^{1/2}.}}
They no longer depend on $f_0$, but still depend on $N_0$ through
factors such as $R_4(\infty)$. 
Given the above data, it is straightforward
to compute $\Sigma$; the answer is
\eqn\dc{\Sigma ={1\over 6}(\sqrt{N_4}+\sqrt{N_{\bar{4}}})^2
(\sqrt{N_5}+\sqrt{N_{\bar{5}}})^2(\sqrt{w_R}+\sqrt{w_L})^2,}
a result similar to \fres.  We have used formulas for $f_i$ at the horizon
which are given by \frel.  
The simple result was not expected
before all other factors magically cancel.
In the extremal limit, $\Sigma$ again is the product 
of numbers of different branes.
This certainly hints at some simple origin.

Another 4D example comes from intersecting 5-branes, 
equation \fourM.  There are
three sets of 5-branes intersecting along $N_1N_2N_3$ 
intersection strings, with gravitational waves travelling
along the the intersections. Taking
the direction of these strings as the longitudinal one, then one
obtains $\Sigma =(1/6)N_1N_2N_3$, exactly the same formula.
In this case, the reduction to IIA 
along $x_{11}$ gives only fourbranes,
\ie\ longitudinal fivebranes, which have a simple description
in matrix theory \grt.
Thus the result \dc\ does not seem so dependent on the particular
configuration of branes, so long as it contains D0-branes.

As in the 5d black hole case, we may define a quantity
$L^6\lambda\sigma_{abs}$ in terms of the absorption cross-section
of supergravitons of wavelength $\lambda$; we have
\eqn\transv{
  L^6\lambda\sigma_{abs}={\pi^3}l_p^9N_1N_2N_3\ ,
}
the same as in \trans\ and agreeing with the transverse volume
up to a trivial numerical factor.

The D0-brane as a probe of the 4d black hole
is different than the scalar considered in
\gk.  Note that even in the extremal limit, the absorption cross section 
computed in \gk\ depends on the longitudinal momentum carried
by the black hole, namely the D0-brane charge. Here again the
probing D0-brane decouples from $N_0$ at the order $v^2$.

\newsec{Matrix black holes}

D-brane technology has enabled a remarkable window into the physics
of the extremal and near-extremal black holes under consideration.
Entropy, Hawking temperature, absorption cross-sections, and greybody
factors all agree with those of macroscopic black holes.
However, the D-brane calculations are valid in the regime $g_{s}Q\ll1$,
while semiclassical black hole physics holds when $g_{s}Q\gg1$,
where $Q$ is a typical charge.  Use of D-brane probes and the
large-N limit may enable a partial bridge of this gap \dps, since 
although the hole is much smaller than the string scale, so is
the size of the probe.  Nevertheless, to address the issue
of Hawking evaporation and quantum coherence, one would like
to directly formulate black hole dynamics in a nonperturbative
framework.  Black holes smaller than the string scale are
not useful in this regard, as one knows that light cones -- 
and therefore horizons -- are fuzzy on the string scale \causality.
Indeed, in the picture of \cm, the information
carried by the black hole resides in the `stringy halo'
surrounding the D-branes.

The matrix model of \bfss\ appears to be such a nonperturbative
formulation in the infinite momentum frame (IMF). 
It consists of matrix quantum mechanics of $N$ D0-branes; the D0-branes
are the partons of the IMF description of eleven-dimensional 
supergravity.  Lorentz invariance and the properties of 
toroidal compactification involve subtleties of the large $N$ limit,
which we will assume can be brought under control.
However, to minimize the effects of any resulting modifications
of our current understanding of the model, it is perhaps best
to choose judiciously
the orientation of the branes in the black hole.
The usual continuum limit is $N\rightarrow\infty$, $R\rightarrow\infty$,
$p_{11}=N/R$ fixed; although there are claims \suss\
that some properties may continue to hold even at finite $N$ and $R$.
Finally, we will use the apparent correspondence between
compactification on a torus, and Yang-Mills
theory on the dual torus \refs{\bfss,\taylor}.

There are several ways one might imagine embedding the black holes
\fiveM\ and \fourM\ into this construction: 

\item{a)}{$x_{11}$ is the coordinate along which the
various branes intersect.  
This is the orientation used in \fiveM\ and \fourM.
In this case, the usual large $N$
limit decompactifies the solution to a 5d or 6d black string.
}

\noindent
Note that in the large $R$ limit, 
the internal wave profile $\varphi_i(u)$
that distinguishes various black holes of given charges becomes 
visible to the asymptotic observer, who can now resolve the
profile using low-energy experiments in the asymptotic region. 
However, if one is merely interested in sending
probes into the black object to learn about Hawking
evaporation, what matters is that there is
a finite capture cross-section for a probe which reaches the
classical horizon in finite proper or affine time coordinate.
The zero-brane partons of the matrix model describe
the eleven-dimensional supergravity multiplet, and 
we have seen that they have these properties.  
A second possibility is to choose (for the 5d black hole)

\item{b)}{$x_{11}$ as a longitudinal coordinate on the
5-brane, not the direction of the intersection string.
}

\noindent
The 2-brane is transverse, and the waves travelling down
the intersection string are not the matrix model partons,
but rather `${\widetilde D0}$ branes' -- carriers of electric
flux in the Yang-Mills theory on the dual torus.
The 2-branes are torons of this Yang-Mills theory \grt.
The $R\rightarrow\infty$ limit again gives a black string.
In this case, the internal waves remain `invisible'
to the macroscopic observer,
since the internal wave profile remains microscopic.
Probes that preserve some supersymmetry
are for example the ${\widetilde D0}$ branes
-- the matrix D0-brane
partons break the supersymmetry of this configuration
(it is not difficult to show that their low-energy
Lagrangian has a static potential).  
Both cases $(a)$ and $(b)$ have the advantage that the 5-branes
involved in the black hole configuration are longitudinal.
A third orientation is

\item{c)}{$x_{11}$ is one of the noncompact coordinates
in which the black hole is a localized object;
the large $N$ limit does not give a black string.
In this case all the 5-branes and 2-branes composing the 
black hole are transverse.}

\noindent
Here there is no problem of principle; the main difficulty is
the lack of a concrete description of the transverse fivebrane,
since it is a magnetic object in the dual Yang-Mills theory.
A similar problem arises in case $(b)$ for the 4d black hole,
where two of the three sets of fivebranes would be transverse.
One can imagine understanding enough of the properties of these objects
to make the same qualitative statements about black hole dynamics
as one can for cases $(a)$ and $(b)$; 
we leave this issue for future research.

In the remainder of this article, we will concentrate on the 
first case -- black strings whose longitudinal coordinate is $x_{11}$ --
as these have perhaps the most straightforward interpretation 
in matrix theory.  An additional advantage is the simple
behavior of the black hole \fiveM, \fourM\ under longitudinal 
boosts.  Finally, one might be able to make contact with
the `transverse volume' measure introduced in section 4.
Since this quantity is independent of zero-brane charge,
it is an $N$-independent, boost-invariant quantity
and therefore is maximally insensitive to any modifications
of the matrix theory which might be required to implement Lorentz
invariance; it might even survive the truncation to finite $N$.

A remarkable picture of probe interaction with a black hole is
developed in \dps\ (see also \dkps).  The static gravitational
field seen by the probe arises from integrating out the
massive open strings that stretch between the probe and the hole.
The resulting moduli space metric is that which should be seen
at long distances and low velocities, in the static coordinates
of an asymptotic observer.  The horizon is a singularity
in the description that appears because the open strings become massless
when the probe reaches the D-brane configuration making up
the black hole.  In D-brane language the horizon is the confluence
of the Coulomb branch of the probe dynamics,
describing separate motion of probe and hole, with the Higgs
branch describing probe-hole bound states.  The vicinity
of this juncture is the so-called `stadium' region \dkps,
where light probe-hole strings contribute important dynamical
effects; in black hole language, one would call this the
`stretched horizon' \stretchor. 

The transcription of the configuration
to matrix theory is straightforward.
In case $(a)$, all membranes and fivebranes are longitudinal.
Longitudinal fivebranes wrapped around a torus are described
in matrix theory as instantons in the dual Yang-Mills theory.
Longitudinal membranes are states
carrying a momentum flux $T_{0i}$ along
the $i^{\rm th}$ internal coordinate \dvv.
The waves along the string are the zero-brane partons themselves.
The 5d black hole turns into a 6d black string; the dual 
Yang-Mills description is a bound state of instantons carrying
momentum on the dual ${\tilde T}^5$.  The 4d black hole becomes
a 5d black string; the corresponding instantons in the 
Yang-Mills theory on the dual ${\tilde T}^6$ are three sets of
two-dimensional objects occupying mutually orthogonal tori.
Shrinking any transverse circle to a size much smaller than
the 11d Planck scale, the dual Yang-Mills theory lives on a
large circle, times a remaining torus of moderate size.
In this limit, one recovers the light-cone description of 
the IIA string \dvv-\imam, and hence in principle the
results of \dps\ (see also \suss).  The matrix description is not
limited to weak string coupling, however.

The translation of the dynamics to matrix theory is as follows:
The D-brane bound state we are describing has a good 
semiclassical limit in the dual Yang-Mills theory as a bound
state carrying various perturbatively visible charges.
The probe is a D0-brane parton (or bound state thereof).
The open strings stretching between the bound state and probe
become off-diagonal elements of the matrices.
As the probe approaches the horizon, it slows down due to 
its interaction with these coordinates.
This is because the D-brane description is intrinsically in 
static coordinates.
When the probe reaches
the black hole `horizon', the off-diagonal entries are
easily excited, and the moduli space metric obtained by integrating
them out becomes singular.  
The full matrix dynamics should be perfectly regular, however.  
The hole-probe system becomes an excited bound state; it can
relax to a BPS state through the emission of low-momentum
partons, which the asymptotic observer interprets as Hawking
radiation.  One expects the spectrum to be thermal, since
the bound state has a long time to explore its phase space
and equilibrate before reradiating.

Why hasn't the information encoded in the infalling probe state
been lost behind an event horizon?  
The basic reason is that the matrix dynamics is framed in
a background Minkowski spacetime, which has no horizons.
What is conventionally thought of as 
the causal structure of spacetime is an effective concept
determined by the moduli space metric
of D0-brane/supergravitons, which breaks down
near the horizon.  One can map out 
this causal structure via the trajectories of
the massless zero-brane test particles, which 
the low-energy observer interprets as lightlike geodesics.
Far from the hole, this provides an accurate picture of information
propagation.  Lightlike geodesics are bent by the `optical medium'
of off-diagonal matrix elements,
mocking up curved space geometry.
These trajectories are dramatically affected 
when the D-brane `stretched horizon' is approached.
The true path taken by such a massless particle involves a
period of thermalization on the stretched horizon,
perhaps followed by reradiation as a Hawking particle
(the Hawking particles may come from other D0-branes present
in the black hole bound state).
Barring a singularity in the large-N physics, it would
appear consistent to interpret the infalling data as getting
stuck on the stretched horizon, thermalized, and reradiated
to the asymptotic region.

\subsec{Crossing the horizon}

How then can one reconstruct the infalling observer's experience?
There ought not to be any such object as a collection of D-branes
that one runs into as one crosses the horizon in finite
proper time.
Therefore, let us examine the change of variables \newcoords\
needed to pass from static to infalling coordinates.  There are
several important features.  

First, the redefinition
$u\rightarrow U=\frac1{2\sigma}e^{2\sigma u}$ brings the horizon
$u\rightarrow\infty$ to a finite coordinate value,
thereby undoing the exponential redshift of infalling
proper time.
There is a nonlocal relation between the matrix 
description and spacetime coordinates.  The IMF description
of the matrix model utilizes $t=x^+=v$ and 
$N/R\sim p_-\sim\frac{\d}{\d x^-}$
as coordinates\foot{Note that our choice of IMF time variable is
just the opposite of that used in, \eg\ \susslorentz, which uses $u$
as the light-cone time of an infalling test string.
Instead, we wish to take $v$ as time, since it is in this
variable that stationary D0-brane probes are BPS saturated.
A small radial D0-brane velocity represents a slight 
departure from BPS, or in other words a lightlike geodesic that
adiabatically crosses the horizon.  Note that this
means that Hawking radiation will be a low-energy
(small longitudinal momentum)
process, rather than a short-time phenomenon at the horizon
as in \susslorentz.}.  
The other light-front coordinate $x^-=u$
is thus dual to $N$.  To localize physics in $x^-$ requires
introduction of a `chemical potential' for $N$, followed
by a Legendre transformation.\foot{We are
only interested in transforming the description of the probe
to its proper time or affine coordinate evolution.  Hence the
appropriate procedure is to consider the
family of probe experiments at different probe
longitudinal momenta $N_{\rm probe}$, and then Legendre transform
this variable to determine the probe wavefunction's 
dependence on $u$.}  Thus the horizon, at $x^-=\infty$ 
as well as $r=0$, is diffused across the probe part of the matrix.

The second important feature is that crossing the horizon at
$r=0$ ($r=r_0$ for a nonextremal hole) is an analytic continuation
of the static radial coordinate to complex values.
The static interior geometry replaces 
$f_i^{\rm ext}=[1+\frac{r_i^2}{r^2}]$ by
$f_i^{\rm ext}=[1-\frac{r_i^2}{r^2}]$ (5d black holes); or
$f_i^{\rm ext}=[1+\frac{r_i}{r}]$ by
$f_i^{\rm ext}=[1-\frac{r_i}{r}]$ (4d black holes).
In both cases, the matrix eigenvalues describing probe physics 
must be continued into the complex domain.

The `internal clock' and other structure of a macroscopic
probe falling into a macroscopic black hole is thus an
approximate construct.  Its evolution inside the horizon is essentially
an analytic continuation -- or rather an extrapolation --
of the probe moduli space approximation (Coulomb branch),
beyond the horizon (juncture with the Higgs branch).
In static coordinates, the probe wavefunction becomes strongly
entangled with that of the black hole, due to their mutual interaction
with the light off-diagonal matrix variables.
The passage to infalling coordinates must represent an approximate
rediagonalization of the matrices, separating probe and black
hole degrees of freedom.
The integrity of the probe wavefunction in these approximate time
and space coordinates can be maintained in 
a limited domain\foot{Much as a heavy object interacting with
a bath of massless objects is not disturbed, so long as the energies
of the light objects does not approach the gap in the excitation
spectrum of the heavy one.  One might then regard the singularity
as the place where this separation fails, due to
high energy processes.}.
At the singularity (or Cauchy horizon, if that is the disease),
the needed transformation becomes singular.
Thus one might regard the infalling description as a kind of
saddle point or collective field
approximation (in complexified matrix space),
which breaks down at singularities of the effective geometry.

It is crucial that the infalling coordinate frame,
and in particular the infalling observer's proper time,
is built out of matrix observables.  Such quantities can be
understood as conventional geometry only when commutators
are small; reaching the singularity (or a Cauchy horizon),
this observer's proper time becomes `noncommutative'.
One can no longer use it to describe a simple semiclassical
evolution equation.  Nevertheless, this is a singularity of
the description, not the physics.  The static (IMF) frame provides
global coordinates for the full evolution,
from infall to evaporation.

This picture of the dynamics may be regarded as a form of
`black hole complementarity' \stretchor-\dvvcomp.
Passage between the static and infalling frame involves
a transform of the matrix variables by left and right
multiplication, and the
two sets of observables will not commute.\foot{For example,
even the instantaneous boost relating a static and infalling
observer near the horizon involves an exponentiation of 
longitudinal boost operators which are built out of 
U(N) generators \dewit.}

\subsec{Fat black holes}

Nonextremal (and even `fat') black holes
can also be described in the matrix framework: 
They are essentially a `plasma' of instantons, anti-instantons,
gluons, etc., of the dual Yang-Mills theory.
The relative amounts of each are controlled by 
the various charges $Q_i\propto N_i-N_{\bar i}$; and by
the shape parameters of the internal torus, 
which act as chemical potentials
by adjusting the masses of the various branes,
\cf\ equations \tsc, \ffs.  This localized lump of plasma
is very long-lived; for example, a graviton attempting
to escape encounters an `optical medium'
whose `refractive index' $n^2=\prod f_i$ has a very strong
radial gradient near the horizon.  If the graviton has any
angular momentum whatsoever, the optics will bend the trajectory
and refocus the escaping wave back onto the hole.
The plasma is supported from indefinite collapse because
gravity is an effective interaction that turns off
at short distances.\foot{G. Polhemus has suggested to us that
eigenvalue repulsion may play an important role
(\cf\ \polhem) in determining the density at which the plasma
stabilizes.}
This picture fits nicely with the transverse volume measure
proposed in section 4.  There it was seen that each
`intersection string' of the constituent branes
occupies on the order of one Planck volume in the
transverse space, for {\it any} black hole
in the class described by \hms.
This suggests that the `plasma' indeed stabilizes 
at Planckian densities.

Finally, it seems that there is a version of
Mach's principle at work here.  There is a preferred frame
imposed on the theory by the underlying Minkowski space
of the Yang-Mills dynamics.  It is the causal structure
of this dynamics, and not the effective trajectories of 
D0-brane/supergravitons, that ensures unitary evolution.
The causal boundaries of the effective gravitational 
physics are failures only of the description of the 
dynamics, and not of the dynamics itself.


\vskip 1cm
\noindent{\bf Acknowledgments} 
We thank 
J. Harvey,
A. Klemm, 
G. Polhemus,
and
S.H. Shenker
for discussions, and
M. Douglas,
G. Horowitz,
and 
D. Marolf
for correspondence.

This work was supported by DOE grant DE-FG02-90ER-40560 and NSF grant
PHY 91-23780.
\listrefs
\end